\documentclass[twocolumn,aps,prb,showpacs,raggedbottom,nobalancelastpage,amssymb,superscriptaddress]{revtex4}

\usepackage{epsfig,amssymb,amsmath,latexsym}
\usepackage{graphics}
\usepackage{subfigure}
\usepackage{wrapfig}
\usepackage{float}
\usepackage{dsfont}
\usepackage[usenames]{color}
\usepackage{soul}
\setulcolor{red}
\setstcolor{red}
\sethlcolor{yellow}

\usepackage{varioref}
\usepackage{subfigure} 
\usepackage{amssymb, amsmath, amsfonts}
\usepackage{pst-plot, pstricks}
\usepackage{setspace}
\usepackage{calc}
\usepackage{array}
\usepackage{layout}
\usepackage{url}
\usepackage{rotating} 
\usepackage{fancybox,amssymb,amsmath}
\usepackage[usenames]{color}
\usepackage{graphicx} 
\usepackage{wasysym}
\usepackage{bbm}
\usepackage{ragged2e} 
\usepackage{hyperref}

\usepackage[title]{appendix}

\begin{document}
\title{Scattering theory of adiabatic reaction forces due to out-of-equilibrium quantum environments}

\author{Mark Thomas}
\author{Torsten Karzig}
\author{Silvia Viola Kusminskiy}
\affiliation{Dahlem Center for Complex Quantum Systems and Fachbereich Physik, Freie Universit\"at Berlin, 14195 Berlin, Germany}
\author{Gergely Zar\'and}
\affiliation{Dahlem Center for Complex Quantum Systems and Fachbereich Physik, Freie Universit\"at Berlin, 14195 Berlin, Germany}
\affiliation{BME-MTA Exotic Quantum Phases ``Lend\"ulet Group'', Institute of Physics, Budapest University of Technology and Economics, H-1521 Budapest, Hungary}
\author{Felix von Oppen}
\affiliation{Dahlem Center for Complex Quantum Systems and Fachbereich Physik, Freie Universit\"at Berlin, 14195 Berlin, Germany}

\date{\today}
\begin{abstract}
The Landauer-B\"uttiker theory of mesoscopic conductors was recently extended to nanoelectromechanical systems. In this extension, the adiabatic reaction forces exerted by the electronic degrees of freedom on the mechanical modes were expressed in terms of the electronic S-matrix and its first non-adiabatic correction, the A-matrix. Here, we provide a more natural and efficient derivation of  these results within the setting and solely with the methods of scattering theory. Our derivation is based on a generic model of a slow classical degree of freedom coupled to a quantum-mechanical scattering system, extending previous work on adiabatic reaction forces for closed quantum systems.

\end{abstract}

\maketitle


\section{Introduction}

The problem of a classical heavy particle embedded in a quantum environment is a paradigm that can be applied to diverse physical systems. The condition for its applicability is the existence of a macroscopic variable that can be treated as classical, coupled to quantum degrees of freedom. If the system allows for a separation of time scales such that the characteristic times of the quantum degrees of freedom are much faster than the classical ones, the evolution can be described within an adiabatic expansion, in which the velocity of the classical variable is taken as a small parameter. The Hamiltonian of the quantum system becomes parametrically dependent on time through the classical degrees of freedom. As the states of the quantum system evolve in time, they acquire a geometric phase, denominated Berry phase, in addition to the usual dynamical phase.\cite{BerryPRSL84} 

The backaction of the quantum environment on the classical degrees of freedom can be cast in terms of effective reaction forces that affect the dynamics of the classical variables. The simplest and best known of these reaction forces is the {\em Born-Oppenheimer} force associated with the adiabatic potential surfaces of the fast quantum system as function of the slow classical variables. The Born-Oppenheimer force depends only on the coordinates of the classical degrees of freedom and is independent of their velocity. As emphasized by Berry\cite{BerryShapWil89} and others\cite{BerryPRSL93,AharonovPRL92,SternPRL92}, additional reaction forces appear when going to next order in the adiabatic approximation, retaining forces which are linear in the velocity of the classical variables. In fact, they found that the Berry phase is mirrored by a Lorentz-like force, which was dubbed ``{\em geometric magnetism}.'' It is not associated to a real magnetic field, but to an emergent geometrical property of the Hilbert space. Like the true magnetic Lorentz force, this emergent Lorentz force is non-dissipative. In general, one may also expect a {\em friction force} in linear order in the velocity of the classical degrees of freedom. However, it was shown by Berry and Robbins \cite{BerryPRSL93} that such a friction force is absent when the quantum system has a discrete spectrum.

\begin{figure}[t]
\begin{centering}
\includegraphics[scale=0.8]{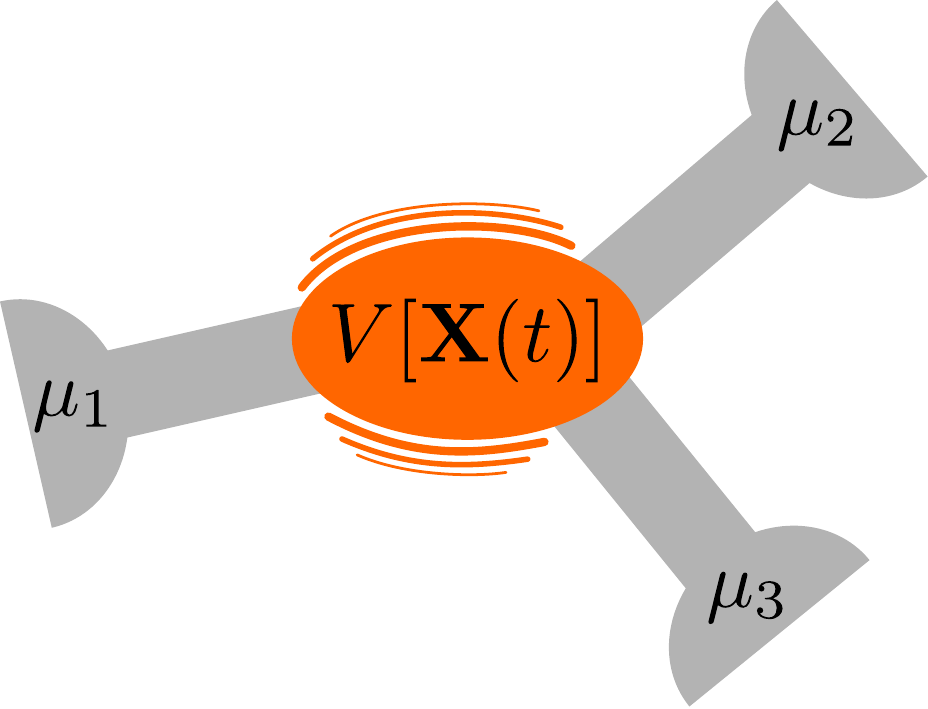}
\end{centering}
\caption{Example of a scattering system that is coupled to slow classical degrees of freedom ${\bf X}(t)$. The movement of the scatterer changes the scattering potential $V[{\bf X}(t)]$. The backaction of the electrons passing through the scatterer then leads to reaction forces acting on ${\bf X}(t)$.} \label{fig:scattering}
\end{figure}

Several recent developments in nanoelectromechanical systems\cite{KindermannPRB02,PistolesiPRB08,LueNanoLett10,BennettPRL10,BodePRL11,BodeBJ12} and spintronics\cite{TserkovnyakPRL02,KupferschmidtPRB06,BrataasPRL08,BrataasPRB11,BodePRB12} suggest to extend these considerations on adiabatic reaction forces to classical degrees of freedom coupled to {\em open} quantum systems out of equilibrium. In this paper, we consider a rather generic model of a quantum mechanical scattering system (such as a coherent mesoscopic conductor within the Landauer-B\"uttiker approach,\cite{LandauerIBM57,LandauerPhilMag70,ButtikerPRB85} see Fig.~\ref{fig:scattering}) which couples to the slow classical system through the scattering potential. Non-equilibrium states of the quantum system can then be modeled by considering different distribution functions for the various incoming scattering channels.\footnote{Note that here, the fast quantum system is open since it is a scattering system coupled to reservoirs. This should be contrasted with the case of an open quantum system involving coupling to a (thermal) bath, usually taken to be an infinite set of harmonic oscillators.\cite{CaldeiraLeggettPRL81} Geometric magnetism in the latter context has recently been considered in Ref.~\onlinecite{CampisiXXX12}.}

Unlike the setting of Berry and Robbins, our scattering-theory setting naturally allows for a friction contribution to the adiabatic reaction forces, even though the fast system is quantum mechanical. By the fluctuation-dissipation theorem, the presence of friction forces requires one to also include a {\em stochastic force}, and the classical degrees of freedom ${\bf X} = \{X_1, X_2,..., X_N\}$ (taken to be mechanical for definiteness) obey a Langevin dynamics,
\begin{equation} 
\label{eq:Langevin}
\dot{P}_\nu -F^{cl}_\nu = F_\nu - \sum_{\nu'} \gamma_{\nu\nu'} \dot{X}_{\nu'} +\xi_\nu\,.
\end{equation}
On the left hand side (LHS), $P_\nu$ denotes the canonical momentum of coordinate $X_\nu$, and we have included the possibility of an external classical force $\mathbf{F}^{cl}({\bf X})$. The adiabatic reaction forces due to the quantum environment are collected on the right hand side (RHS) of  Eq.~\eqref{eq:Langevin}, where $\mathbf{F}({\bf X})$ is the Born-Oppenheimer force exerted by the environment, while $\mathbf{\xi}$ denotes the stochastic Langevin force which represents fluctuations on top of $\mathbf{F}$. The dissipative and Lorentz-like forces are encoded in the symmetric and antisymmetric parts, respectively, of the tensor $\boldsymbol{\gamma}({\bf X})$.

The scattering approach suggests that all adiabatic reaction forces can be expressed in terms of the S matrix (including non-adiabatic corrections) of the quantum system. These expressions were obtained in previous work,\cite{BodePRL11,BodeBJ12} based on a Keldysh Green's function approach for a closely related model. Here, our primary aim is to derive these expressions [given in Eqs.~\eqref{eq:Born-Oppenheimer_force},~\eqref{eq:sym_friction},~\eqref{eq:antisym_friction}, and \eqref{eq:result_correlator}] directly within the setting and with methods of scattering theory. This alternative derivation has several advantages: (i)  In avoiding Keldysh Green's functions extraneous to scattering theory, the derivation is both more natural and more direct. (ii) The generic scattering theory formulation emphasizes the generality and broad applicability of the results. (iii) The approach also brings out similarities with and differences  from the seminal considerations of Berry and Robbins\cite{BerryShapWil89,BerryPRSL93} for closed quantum systems. 

While our model and our results are quite generic, a key motivation was provided by nanoelectromechanical systems\cite{BunchScience07,GalperinJoPCM07,LassagneScience09,SteeleScience09,GanzhornPRL12} and spintronics\cite{FertRMP08,RalphJoMMM08,MisiornyPSS09} devices. In these systems, the motion of the mechanical mode or the localized spin can frequently be thought of as a slow classical degree of freedom while the electronic conduction is quantum coherent and can be described as a quantum-mechanical scattering system, following  Landauer and B\"uttiker. An important focus of recent work on adiabatic reaction forces in  nanoelectromechanical and spintronics systems, often termed current-induced forces in this context, are the qualitatively new features introduced by out-of-equilibrium quantum environments. It is now well understood\cite{ClerkNJP05,KuznetsovJPCM08,HusseinPRB10,TodorovPRB10,LueNanoLett10,BodePRL11,LuePRL11,BodeBJ12,LuePRB12} that for non-equilibrium environments, (i) the Born-Oppenheimer force is in general no longer conservative and thus cannot be obtained from a potential surface; (ii) it is possible to have negative dissipation; and (iii) a Lorentz-like force can emerge even for time-reversal invariant conductors. Of course, the approach taken here reproduces all of these results. 

Our approach may have other interesting applications. Since it can be applied similarly to both fermionic and bosonic environments it also provides a scattering description of adiabatic reaction forces in optomechanical \cite{KippenbergSci08} or cold-atom systems.\cite{DalibardRMP11} Moreover it is also interesting to compare them to older results on the motion of vortices in superfluids. There it was shown that the geometric Berry phase is responsible for the Magnus force on a vortex.\cite{AoPRL93} Later it was also realized that dissipation can be obtained in an analogous manner within an adiabatic expansion, by allowing for broadening of the energy levels of the system.\cite{GaitanPRA98,ZhuJLTP98} This broadening stems from the connection of the systems to an environment which is naturally implemented in our scattering approach.

This manuscript is organized as follows. In Sec. \ref{sec:ScattTheo} we present the basic tools of scattering theory that are needed for the derivation of the adiabatic reaction forces, and we express the adiabatic expansion of the S-matrix in terms of frozen scattering states. In Sec. \ref{sec:Forces} we derive expressions for the adiabatic reaction forces appearing on the RHS of the Langevin equation in Eq.~\eqref{eq:Langevin}, in terms of the S-matrix of the quantum mechanical scattering system, including the first non-adiabatic correction. We conclude in Sec. \ref{sec:Concl}. We relegated some details to App.~\ref{sec:AandS-matrix}, and connect our results to those found in Refs.~\onlinecite{BodePRL11} and \onlinecite{BodeBJ12} in App.~\ref{sec:Application}.

\section{Scattering theory and adiabatic expansion}
\label{sec:ScattTheo}
\subsection{Elements of Scattering Theory}
In this section we introduce necessary aspects of scattering theory. We consider a system described by a single-particle Hamiltonian $H= H_0 + V$, where $H_0$ is a free Hamiltonian and $V$ is a scattering potential which depends parametrically on time through the slowly varying classical degrees of freedom $\mathbf{X}(t)$. In order to describe the system in terms of scattering states, $V$ is assumed to be confined to a finite region in space. The time dependent retarded ($+$) and advanced ($-$) scattering states $| \Psi_m^{\pm} (\varepsilon,t) \rangle$ are solutions of the \emph{time dependent} Schr\"odinger equation (note that throughout this work we set $\hbar=1$)
\begin{equation}\label{eq:Schroedinger}
 i\,\partial_t \big| \Psi_m^{\pm} (\varepsilon,t) \big\rangle =   H\, \big| \Psi_m^{\pm} (\varepsilon,t) \big\rangle\,,
\end{equation}
where the index $m$ is a combined index labeling channels and leads. It is convenient to define scattering states $| \psi_m^{\pm}(\varepsilon,t) \rangle $ \emph{without} the dynamical phase, 
\begin{equation}\label{eq:PsiChi}
\big| \Psi_m^{\pm} (\varepsilon,t) \big\rangle =  \mathrm{e}^{-i\,\varepsilon\,t} \big| \psi_m^{\pm}(\varepsilon,t) \big\rangle\,, 
\end{equation} 
which fulfill 
\begin{equation}\label{eq:ModSchroedinger}
 i\,\partial_t \big| \psi_m^{\pm} (\varepsilon,t) \big\rangle =   \left(H-\varepsilon\right)\, \big| \psi_m^{\pm} (\varepsilon,t) \big\rangle\,.
\end{equation}
The advanced and retarded scattering states are specified through their boundary conditions. While the retarded state $|\psi_m^+\rangle$ has incoming waves only in channel $m$, the advanced state $|\psi_m^-\rangle$ has outgoing waves only in this channel,     
\begin{equation}\label{eq:free_sc_st}
\big| \psi_m^{\pm}(\varepsilon,t\rightarrow\mp\infty) \big\rangle= \big| \phi_m (\varepsilon) \big\rangle \,,
\end{equation}
where $| \phi_m (\varepsilon) \rangle$ is the eigenstate of the free Hamiltonian in channel $m$, 
\begin{equation}\label{eq:freestates}
 H_0 \big| \phi_m (\varepsilon) \big\rangle =   \varepsilon \big|  \phi_m (\varepsilon) \big\rangle\,.
\end{equation}
Eq. \eqref{eq:free_sc_st} holds in the weak sense, {\it i.e.}, wave packets constructed from the scattering states $| \Psi_m^{\pm} (\varepsilon,t)\rangle$ behave as free wave packets for times $t\to \pm \infty $, and have energy $\varepsilon$. We normalize the scattering states such that $|\phi_m\rangle$ has unit flux, which implies the orthonormality relations 
\begin{equation}\label{eq:normalization}
\begin{split}
\big\langle \psi_m^{\pm} \left(\varepsilon,t\right) \big| \psi_{m^\prime}^{\pm} \left(\varepsilon^\prime,t\right) \big\rangle&=
\big\langle \phi_m \left(\varepsilon\right) \big| \phi_{m^\prime} \left(\varepsilon^\prime\right) \big\rangle\\
&=2\pi\delta\left(\varepsilon-\varepsilon^\prime\right)\delta_{mm^\prime}\,.
\end{split}
\end{equation}

In the strictly adiabatic limit, the wave function $| \psi_m^{\pm} (\varepsilon,t) \rangle$ is time-independent and hence Eq.~\eqref{eq:ModSchroedinger} reduces to the time-independent Schr\"odinger equation
\begin{equation}\label{eq:adscst}
 H_t\big| \psi_m^{{\bf X}_t\pm} (\varepsilon) \big\rangle =  \varepsilon \big|  \psi_m^{{\bf X}_t\pm} (\varepsilon) \big\rangle
\end{equation}
for a frozen configuration of the potential $V_t=V(\mathbf{X}_t)$, where ${\bf X}_t={\bf X}(t)$ and we have also defined $H_t=H(\mathbf{X}_t)$. We denote the frozen scattering states by $| \psi_m^{{\bf X}_t\pm} (\varepsilon) \rangle$. The superscript ${\bf X}_t$ or subscript $t$ emphasizes the parametric dependence on time of each quantity due to the slow variation of the scattering potential. Introducing frozen Green's functions 
\begin{equation}\label{eq:GRA}
G_t^{R/A}(\varepsilon) = \frac{1}{(\varepsilon-H_t\pm i\eta)} \,,
\end{equation}
($\eta\rightarrow 0^+$) we can write the Lippmann-Schwinger equation for the frozen scattering states,
\begin{equation}\label{eq:LippmanSchwinger}
\big|\psi^{{\bf X}_t\pm}_m(\varepsilon)\big\rangle = \big|\phi_m^{\vphantom{t}}(\varepsilon)\big\rangle + G_t^{R/A}(\varepsilon) \,V_t\,\big|\phi_m^{\vphantom{t}}(\varepsilon)\big\rangle\,,
\end{equation}
where $ |\phi(\varepsilon)\rangle$ are the free eigenstates introduced in Eq.~\eqref{eq:freestates}. Eq.~\eqref{eq:LippmanSchwinger} will be of use in the next subsection.

The frozen S-matrix $S_t(\varepsilon)$ is defined by the overlap of the frozen retarded and advanced scattering states, and hence depends only on the energy $\varepsilon$ of the incoming states,
\begin{equation}\label{eq:FrozenS}
S _t^{n k}(\varepsilon)2\pi \delta(\varepsilon-\varepsilon') = \big\langle \psi^{{\bf X}_t-}_n (\varepsilon')  \big| \psi^{{\bf X}_t+}_k (\varepsilon) \big\rangle\,,
\end{equation}
where we have isolated the singular dependence on energy, $\delta(\varepsilon-\varepsilon')$. The frozen S-matrix is unitary, $S_t(\varepsilon) S_t(\varepsilon)^\dagger = \mathbbm{1}$, since scattering states are assumed to be normalized to unit flux. For a slowly changing system, the frozen S-matrix is the zeroth order contribution to the full S-matrix in an adiabatic expansion, as we show in the next subsection, and it depends parametrically on time through the slowly varying parameters $\mathbf{X}_t$.

The \emph{exact} scattering matrix for the time-dependent problem is defined by the overlap of the time-dependent scattering states introduced in Eq.~\eqref{eq:Schroedinger},
\begin{equation}\label{eq:full_S}
 \mathcal{S}_{n k}(\varepsilon',\varepsilon) = \big\langle \Psi_n^{-} (\varepsilon',t_0)  \big| \Psi_k^{+} (\varepsilon,t_0) \big\rangle\,.
\end{equation}
The exact scattering matrix is also unitary due to the unit flux normalization condition 
\begin{equation}\label{eq:UnitaryS}
\sum_n\int\frac{ d \varepsilon}{2\pi}\mathcal{S}_{mn}(\varepsilon',\varepsilon)\mathcal{S}_{nk}\dagger(\varepsilon,\varepsilon'')=2\pi\delta(\varepsilon'-\varepsilon'')\delta_{mk}\,.
\end{equation}
It is important to note that the time $t_0$ at which the overlap of the scattering states is evaluated in Eq.~\eqref{eq:full_S}, can be chosen arbitrarily. The independence of $t_0$ can be seen by taking the derivative with respect to time of Eq.~\eqref{eq:full_S}, and using the Schr\"odinger equation \eqref{eq:Schroedinger}. This allows us to choose $t_0$ in a convenient manner in the following section.

We will see in Sec.~\ref{sec:Forces} that even for a slow evolution, corrections to the adiabatic solution are important to describe the environment-induced forces. Hence we devote the next subsection to calculating the first non-adiabatic correction to the frozen S-matrix.

\subsection{Adiabatic expansion and A-matrix}

The adiabatic expansion relies on the assumption that the classical degrees of freedom ${\bf X}_t$ vary slowly in time. We characterize this slow time dependence by a typical frequency $\Omega$.  In  finite quantum systems, adiabaticity requires $\Omega$ to be small compared to the level spacing $\Delta$. This condition is obviously violated in the open quantum systems of interest here, which have a continuous spectrum. For these systems, adiabaticity requires $\Omega$ to be small compared to the inverse dwell time of the electrons in the scattering region,\cite{WignerPR55} $\Omega \ll 1/\tau_D$. 

The adiabatic expansion is conveniently carried out in the Wigner representation
\begin{equation}\label{eq:WTfull_S}
\mathcal{S}\left(\varepsilon,t\right)=\int\frac{\mathrm{d}\tilde{\varepsilon}}{2\pi}\,\mathrm{e}^{-i\tilde{\varepsilon}t}\,\mathcal{S}\left(\varepsilon+\tilde{\varepsilon}/2,\varepsilon-\tilde{\varepsilon}/2\right)
\end{equation}
of the full S-matrix ${\cal S}(\varepsilon',\varepsilon)$. In the adiabatic limit, the S-matrix depends only slowly on the central time $t$.  In fact, in the limit of a static Hamiltonian, $S(\varepsilon, t)$  becomes independent of $t$ and reduces to the frozen S-matrix $S_t(\varepsilon)$.

For a slowly time-dependent scattering potential the exact S-matrix $\mathcal{S}$ can be written, up to first order in the adiabatic expansion, as~\cite{MoskaletsPRB04,MoskaletsPRB05,BodePRL11}
\begin{equation}\label{eq:S_expand}
\mathcal{S}(\varepsilon,t)= S_t(\varepsilon)+A_t(\varepsilon)+\ldots\,,
\end{equation}
where all quantities on the RHS depend parametrically on time. Eq.~\eqref{eq:S_expand} defines the A-matrix
\begin{equation}\label{eq:Aalpha}
A_t(\varepsilon)=\sum_{\nu=1}^N A_t^\nu(\varepsilon)\dot{X}_\nu
\end{equation}
as the first-order non-adiabatic correction of $\mathcal{S}(\varepsilon,t)$, which depends linearly on the velocity $\dot{\mathbf{X}}$ and parametrically on time through $\mathbf{X}(t)$ . Below, we derive an important relation which expresses the A-matrix in terms of the frozen scattering states $| \psi^{{\bf X}_t\pm} (\varepsilon) \rangle$,
\begin{equation}\label{eq:A}
\begin{split}
 A_t^{\nu,{nk}}(\varepsilon)&= \frac{1}{2}\big\langle \partial_{\varepsilon}\psi_n^{{\bf X}_t-}\big|\partial_\nu V_t\big| \psi_k^{{\bf X}_t+} \big\rangle\\
& - \frac{1}{2}\big\langle \psi_n^{{\bf X}_t-}\big|\partial_\nu V_t\big|\partial_{\varepsilon} \psi_k^{{\bf X}_t+} \big\rangle\,,
\end{split}
\end{equation} 
where $\partial_\nu=\partial/\partial X_\nu$ and $|\partial_{\varepsilon} \psi_k^{{\bf X}_t+}\rangle=\partial_{\varepsilon} |\psi_k^{{\bf X}_t+}\rangle $. In previous works the A-matrix was given in terms of Green's function expressions, \cite{BodePRL11,BodeBJ12,VavilovPRB01,ArracheaPRB06} or obtained by expanding the exact solution of the time dependent problem.\cite{MoskaletsPRB04,MoskaletsPRB05} Equation~\eqref{eq:A} provides a systematic way of obtaining $A$ from the solution of the static scattering problem.

To derive Eq.~\eqref{eq:A}, we first compute the scattering states $|\psi^\pm (\varepsilon,t)\rangle$ to first order in the adiabatic expansion,\cite{Entin-WohlmanPRB02}  
\begin{equation}\label{eq:expand}
\big|\psi^{\pm}(\varepsilon,t) \big\rangle= \big|\psi^{{\bf X}_t\pm}(\varepsilon)\big\rangle+\big|\delta\psi^{{\bf X}_t\pm}(\varepsilon)\big\rangle+\ldots\,.
\end{equation}
Here, the frozen scattering state $|\psi^{{\bf X}_t\pm}\rangle$ is the zeroth-order term in the adiabatic expansion and corresponds to the strictly adiabatic limit, while $|\delta\psi^{{\bf X}_t\pm}\rangle $ denotes the first non-adiabatic correction~\footnote{In Eq. \eqref{eq:expand}, an overall phase may in principle also appear on the R.H.S.. However this phase does not influence our final results since its time evolution is taken into account through the adiabatic expansion.}. (Here, we omit the channel index for notational simplicity.) Inserting Eq.~\eqref{eq:expand} into the Schr\"odinger equation Eq.~\eqref{eq:ModSchroedinger}, using Eq.~\eqref{eq:adscst}, and comparing terms of first order in the adiabatic expansion, we find 
\begin{equation}\label{eq:chiexp}
i\partial_t  \big|\psi^{{\bf X}_t\pm}(\varepsilon)\big\rangle=(H_t-\varepsilon)\big|\delta\psi^{{\bf X}_t\pm}(\varepsilon)\big\rangle\,,
\end{equation}
where $\partial_t$ indicates the parametric derivative with respect to time. With Eq. \eqref{eq:GRA}, we conclude that $|\delta\psi^{{\bf X}_t\pm}\rangle=-i\,G^{R/A}_t\partial_t|\psi^{{\bf X}_t\pm}\rangle$ and hence, plugging this back into Eq.~\eqref{eq:expand},
 \begin{equation}\label{eq:adiabatic_state_expansion}
 \big|\psi^{\pm}(\varepsilon,t)\big\rangle=\big|\psi^{{\bf X}_t\pm}(\varepsilon)\big\rangle-i\,G^{R/A}_t(\varepsilon)\,\partial_t\big|\psi^{{\bf X}_t\pm}(\varepsilon)\big\rangle+\ldots\,.
\end{equation}

We can also express $\partial_t|\psi^{{\bf X}_t\pm}\rangle$ in terms of the frozen scattering states by using the Lippmann-Schwinger equation given in Eq.~\eqref{eq:LippmanSchwinger}. Taking the time derivative of Eq.~\eqref{eq:LippmanSchwinger} and using that $\dot{G}_t^{R/A}=G^{R/A}_t\dot{V}_tG^{R/A}_t$, where $\dot{V}_t=\partial_t V_t $, the time derivative of the scattering states can be expressed as \cite{Entin-WohlmanPRB02} 
\begin{equation}\label{eq:derivative}
  \partial_t\big|\psi^{{\bf X}_t\pm}(\varepsilon) \big\rangle = G^{R/A}_t(\varepsilon) \dot{V}_t\,\big|\psi^{{\bf X}_t\pm}(\varepsilon) \big\rangle\,,
\end{equation}
and hence we obtain the desired result from Eq.~\eqref{eq:adiabatic_state_expansion},
\begin{equation}\label{eq:scatteringstates}
\big|\psi^{\pm}(\varepsilon,t)\big\rangle=\big|\psi^{{\bf X}_t\pm}(\varepsilon)\big\rangle-i\left(G^{R/A}_t\right)^{2}\dot{V}_t\big|\psi^{{\bf X}_t\pm}(\varepsilon)\big\rangle+\ldots\,,
\end{equation}
which is valid to first order in the adiabatic expansion.

With this expansion of the scattering states, the adiabatic expansion of the full S-matrix can be performed, starting from the definition Eq.~\eqref{eq:full_S} and the Wigner transform given in Eq.~\eqref{eq:WTfull_S}. It is convenient to use the arbitrariness of $t_0$ in Eq.~\eqref{eq:full_S} by choosing $t_0$ as the central time, $t_0=t$. Defining $\varepsilon_\pm=\varepsilon\pm\tilde{\varepsilon}/2$, the Wigner transformed S-matrix can now be approximated to first order in the adiabatic expansion as
\begin{equation}
\begin{split}
\mathcal{S}(\varepsilon,t)&=\int\frac{\mathrm{d}\tilde{\varepsilon}}{2\pi}\mathrm{e}^{-i\tilde{\varepsilon}t}\big\langle \Psi^{-}\left(\varepsilon_+,t\right)\big|\Psi^{+}\left(\varepsilon_-,t\right)\big\rangle \\
&= \int\frac{\mathrm{d}\tilde{\varepsilon}}{2\pi}\big\langle \psi^{{\bf X}_t-}\left(\varepsilon_+\right)\big|\psi^{{\bf X}_t+}\left(\varepsilon_-\right)\big\rangle\\
&-i\int\frac{\mathrm{d}\tilde{\varepsilon}}{2\pi}\big\langle \psi^{{\bf X}_t-}\left(\varepsilon_+\right)\big|\left[G^{R}_t(\varepsilon_-)\right]^{2}\dot{V}_t\big|\psi^{{\bf X}_t+}\left(\varepsilon_-\right)\big\rangle\\
&+i\int\frac{\mathrm{d}\tilde{\varepsilon}}{2\pi}\big\langle \psi^{{\bf X}_t-}\left(\varepsilon_+\right)\big|\dot{V}_t\left[ G^{R}_t(\varepsilon_+)\right]^{2}\big|\psi^{{\bf X}_t+}\left(\varepsilon_-\right)\big\rangle\\
&+ \ldots\,,
\end{split}
\end{equation}
where we have used Eq.~\eqref{eq:scatteringstates}. We now employ the identities $G^{R}_t(\varepsilon_\pm)|\psi^{{\bf X}_t+}\left(\varepsilon_\mp\right)\rangle=(\pm\tilde{\varepsilon}+i\eta)^{-1}|\psi^{{\bf X}_t+}\left(\varepsilon_\mp\right)\rangle$ and $\left[(\tilde{\varepsilon}+i\eta)^{-2}-(-\tilde{\varepsilon}+i\eta)^{-2}\right]=2\pi i\partial_{\tilde{\varepsilon}} \delta(\tilde{\varepsilon})$ to obtain
\begin{equation}
\begin{split}
\mathcal{S}(\varepsilon,t)&=  S_t(\varepsilon)\\
&-\int\mathrm{d}\tilde{\varepsilon}\left[\partial_{\tilde{\varepsilon}}\delta(\tilde{\varepsilon})\right]\big\langle \psi^{{\bf X}_t-}\left(\varepsilon_+\right)\big|\dot{V}_t\big|\psi^{{\bf X}_t+}\left(\varepsilon_-\right)\big\rangle+\ldots\, .
\end{split}
\end{equation}
Integrating by parts with respect to $\tilde{\varepsilon}$, we find
\begin{equation}
\begin{split}
\mathcal{S}(\varepsilon,t)  &= S_t(\varepsilon)+\frac{1}{2}\big\langle \partial_{\varepsilon}\psi^{{\bf X}_t-}\left(\varepsilon\right)\big|\dot{V}_t\big|\psi^{{\bf X}_t+}\left(\varepsilon\right)\big\rangle \\
&-\frac{1}{2}\big\langle \psi^{{\bf X}_t-}\left(\varepsilon\right)\big|\dot{V}_t\big|\partial_{\varepsilon}\psi^{{\bf X}_t+}\left(\varepsilon\right)\big\rangle+\ldots\,, 
\end{split}
\end{equation}
which gives the full S-matrix $\mathcal{S}$ in terms of the frozen S-matrix $S_t$ defined in Eq.~\eqref{eq:FrozenS}, and the first non-adiabatic correction matrix $A_t$ (A-matrix) as anticipated in Eq.~\eqref{eq:A}. \footnote{Eq.~\eqref{eq:A} is closely related to the first non-adiabatic correction considered in Ref.~\onlinecite{Entin-WohlmanPRB02} in terms of scattering states. However we find that the expressions involving terms of the type $\langle\psi^+|\partial_t{\psi^-}\rangle$ which are given in Ref.~\onlinecite{Entin-WohlmanPRB02} are ill defined, giving rise to divergences when applied to simple, analytically solvable scatterers involving, e.g., $\delta$-function potentials.} 

We finish this section by deriving some identities for the S and A-matrices that will be of use in the derivation of the adiabatic reaction forces. The frozen S-matrix can be written as 
\begin{equation}\label{eq:STmat}
S_t^{n k} (\varepsilon) = \delta_{n k} -  i \big\langle \psi_n^{{\bf X}_t-}(\varepsilon)\big|\, V_t \,\big| \phi_k(\varepsilon) \big\rangle\,,
\end{equation}
which follows from recognizing that the second term on the RHS of~\eqref{eq:STmat} is the frozen $T$-matrix.\cite{Roman65} This together with Eqs.~\eqref{eq:LippmanSchwinger} and~\eqref{eq:derivative} gives the time derivative of the frozen S-matrix in terms of the frozen scattering states,
 \begin{equation}\label{eq:dtS}
  \partial_t S_t^{n k} (\varepsilon) = -  i \, \big\langle \psi_n^{{\bf X}_t-}(\varepsilon)\big|\,\dot{V}_t \,\big| \psi_k^{{\bf X}_t+}(\varepsilon) \big\rangle \,.
 \end{equation}
Hence
\begin{equation}
\partial_{\varepsilon}\big\langle \psi^{{\bf X}_t-}\left(\varepsilon\right)\big|\dot{V}_t\big|\psi^{{\bf X}_t+}\left(\varepsilon\right)\big\rangle =i\partial_{\varepsilon}\partial_{t}S_t(\varepsilon) 
\end{equation}
and we obtain an alternative expression for the A-matrix by comparing with Eq.~\eqref{eq:A},
\begin{equation}\label{eq:A2}
A_t(\varepsilon)=-\big\langle \psi^{{\bf X}_t-}\left(\varepsilon\right)\big|\dot{V}_t\big|\partial_{\varepsilon}\psi^{{\bf X}_t+}\left(\varepsilon\right)\big\rangle +\frac{i}{2}\partial_{\varepsilon}\partial_{t}S_t(\varepsilon)\,.
\end{equation}

The S and A-matrices are related through unitarity of the exact S-matrix, \cite{MoskaletsPRB04,MoskaletsPRB05,BodePRL11,BodeBJ12} resulting in the identity
\begin{equation}\label{eq:S_and_A}
 S_t^{\dagger}A_t+A_t^{\dagger}S_t=\frac{i}{2}\left(\partial_{t}S_t^{\dagger}\partial_{\varepsilon}S_t-\partial_{\varepsilon}S_t^{\dagger}\partial_{t}S_t\right)\,.
 \end{equation}
where all quantities are evaluated at the same energy. We can check that our explicit expression for the A-matrix in Eq.~\eqref{eq:A} indeed fulfills this condition. This is shown in App.~\ref{sec:AandS-matrix}.

\section{Adiabatic Reaction Forces}
\label{sec:Forces}
\subsection{Adiabatic reaction forces and scattering states}
The force operator in the Schr\"odinger picture can be defined as 
\begin{equation}
\hat {\mathcal{F}}_{\bf X} = -  \nabla\mathcal{H}\,,
\label{force_op}
\end{equation}
where the gradient is taken with respect to $\mathbf{X}$ and $\mathcal{H} = \mathcal{H}({\bf X} )$
is the (non-interacting) many-body Hamiltonian of the quantum system. 
Notice that 
$\mathcal{H}$ includes terms arising from the free Hamiltonian $H_0$ of the fast degrees of freedom and the scattering potential $V({\bf X})$ which depends parametrically on the slow, classical variables $\mathbf{X}$.

Then for a given trajectory ${\bf X}_t$ the average force that the out-of-equilibrium quantum environment exerts on the classical degrees of freedom ${\bf X}$ at time $t$ is given by
\begin{equation}
\label{eq:FnablaH}
\mathcal{F} (t) = \mathcal{F} [{\bf X}_t] = \langle 
\hat {\mathcal{F}}_{{\bf X}_t}
\rangle\,.
\end{equation}
Here the expectation value indicates quantum-statistical averaging for a given trajectory $\mathbf{X}_t$:  $\langle\ldots\rangle= \rm{Tr}\{\rho(t)\ldots\}$, where $\rho(t)$ is the many-body density matrix of the system at time $t$.  
Notice that $\rho(t)$ and thus the force $\mathcal{F}$ are  {\em functionals} of 
the trajectory  $\mathbf{X}_t$, and therefore $\mathcal{F}$ depends on time through $\mathbf{X}_t$ and its time derivatives, 
$\dot{\mathbf{X}}_t$, $\ddot{{\mathbf{X}}}_t, \dots$.  In fact, the adiabatic expansion consists of making a systematic  expansion in these latter quantities. 
Also, equation~\eqref{eq:FnablaH} gives only the average force. As mentioned in the Introduction, the Langevin dynamics includes the stochastic fluctuations of $-\nabla {\mathcal H}$, which we will consider further below.

To compute the quantum-statistical average $\langle \ldots \rangle$ in Eq.~\eqref{eq:FnablaH}, we write the many-body Hamiltonian in  terms of creation and annihilation operators $a_n^\dagger(\varepsilon,t)$ and $a_n(\varepsilon,t)$ that create/annihilate the retarded scattering states $|\Psi_n^+(\varepsilon,t)\rangle$. Since the time evolution is unitary, the retarded scattering states constitute a complete basis at any time $t$. Note that we are working in the Schr\"odinger representation, and the time $t$ appears as a label in the creation/annihilation operators $a$ simply to identify the corresponding basis. Hence we have 
\begin{equation}
\mathcal{H}_t=\int\frac{\mathrm{d}\varepsilon}{2\pi}\int\frac{\mathrm{d}\varepsilon'}{2\pi}\sum_{mk}\left[H_t\right]_{mk}a_{m}^{\dagger}(\varepsilon,t)a_{k}(\varepsilon',t).
\end{equation}
It is straightforward to show that the quantity $\big\langle a_{m}^{\dagger}(\varepsilon,t)a_{k}(\varepsilon',t)\big\rangle$ is independent of time, by noting that both the retarded scattering states $|\Psi_n^+(\varepsilon,t)\rangle$ and the density matrix $\rho$ evolved unitarily from the unperturbed states. Then, the occupation $f_n(\varepsilon)$ of a scattering state in channel $n$ is governed by the corresponding reservoir, as in the Landauer-B\"uttiker theory of mesoscopic conductors,
\begin{equation}
 \big\langle a_{m}^{\dagger}(\varepsilon,t)a_{k}(\varepsilon',t)\big\rangle =f_{k}(\varepsilon)\delta_{km}2\pi\delta(\varepsilon-\varepsilon')\,.
\end{equation}
Expressing the force operator $\hat {\cal F}_{{\bf X}_t}= -\nabla {\mathcal H}_t$ in terms of these creation and annihilation operators as
\begin{equation}
\mathcal{\nabla H}_t=\int\frac{\mathrm{d}\varepsilon}{2\pi}\int\frac{\mathrm{d}\varepsilon'}{2\pi}\sum_{mk}\left[\nabla H_t\right]_{mk}a_{m}^{\dagger}(\varepsilon,t)a_{k}(\varepsilon',t),
\end{equation}
we are now in a position to evaluate the average adiabatic reaction force as
\begin{equation}\label{DefFullF}
\mathcal{F}=-\sum_{k}\int\frac{\mathrm{d}\varepsilon}{2\pi}f_{k}(\varepsilon)\big\langle \psi_{k}^{+}(\varepsilon,t)\big|\nabla H_t^{\vphantom{+}}\big|\psi_{k}^{+}(\varepsilon,t)\big\rangle \,.
\end{equation}
This expression allows us to perform an adiabatic expansion of the reaction force using the adiabatic expansion of the scattering states developed in Sec.~\ref{sec:ScattTheo}.
 
Inserting the adiabatic expansion of the scattering states given by Eq.~\eqref{eq:scatteringstates} into Eq.~\eqref{DefFullF}, and keeping terms up to first order in the adiabatic expansion, we find
\begin{equation}\label{eq:FAdExp}
\begin{split}
\mathcal{F}[\mathbf{X}_t]&=-\int\frac{\mathrm{d}\varepsilon}{2\pi}f_k(\varepsilon)\big\langle \psi_k^{{\bf X}_t+}(\varepsilon)\big|\nabla V_t\big|\psi_k^{{\bf X}_t+}(\varepsilon)\big\rangle\\
&-i\int\frac{\mathrm{d}\varepsilon}{2\pi}f_k(\varepsilon)\big\langle \psi_k^{{\bf X}_t+}(\varepsilon)\big|\partial_t{V}_t\left(G^{A}_t\right)^{2}\nabla V_t\big|\psi_k^{{\bf X}_t+}(\varepsilon)\big\rangle\\
&+i\int\frac{\mathrm{d}\varepsilon}{2\pi}f_k(\varepsilon)\big\langle \psi_k^{{\bf X}_t+}(\varepsilon)\big|\nabla V_t\left(G^{R}_t\right)^{2}\partial_t{V}_t\big|\psi_k^{{\bf X}_t+}(\varepsilon)\big\rangle \\
&+\ldots\,,
\end{split}
\end{equation}
where we have used that $\nabla H_t=\nabla V_t$ and left the summation over the channel index $k$ implicit.

Equation~\eqref{eq:FAdExp} yields the deterministic reaction forces appearing on the RHS of the Langevin equation, Eq.~\eqref{eq:Langevin}. The zeroth order term, given by the first line of Eq.~\eqref{eq:FAdExp}, is independent of the velocity $\dot {\bf X}$ and gives the (possibly non-conservative) Born-Oppenheimer force $\mathbf{F}(\mathbf{X})$. The first-order contribution, given by the second and third terms of Eq.~\eqref{eq:FAdExp}, represents the forces that depend linearly on the velocity of the classical modes,  $-{\boldsymbol\gamma}(\mathbf{X})\cdot\dot{\mathbf{X}}$.

Let us now turn to the force fluctuations. To define time-dependent 
force fluctuations at the quantum mechanical level, one needs to go to the Heisenberg picture, 
$\hat {\mathcal{F}}_{\bf X} \to \hat {\mathcal{F}}_{\bf X}(t)$ and define the Heisenberg force fluctuation 
operators, 
\begin{equation}\label{eq:DefXi}
\hat{\mathbf{\xi}} (t) \equiv \hat {\mathcal{F}}_{{\bf X}_t}(t) - \mathcal{F} (t).
\end{equation}
There are two different contributions to the stochastic force: (i) fluctuations at finite temperatures and (ii) non-equilibrium noise which is a consequence of the probabilistic nature of the scattering process. Since the quantum degrees of freedom are fast compared to the mechanical motion, the correlator $D_{\alpha\beta}(t,t')$ of the stochastic force is local on the relevant time scales of the Langevin equation \eqref{eq:Langevin},
\begin{equation}\label{eq:D_alpha_beta}
D_{\alpha\beta}(t,t')= \big\{\big\langle\hat{\xi}_{\alpha}(t)\hat{\xi}_{\beta}(t')\big\rangle\big\}_{s}\simeq D_{\alpha\beta}(t)\delta(t-t')\,,
\end{equation}
were the subscript $s$ denotes symmetrizing with respect to $\alpha$ and $\beta$.
To account for these fluctuations and to satisfy the fluctuation-dissipation theorem for the classical variable $\bf X$, 
one must incorporate  in Eq.~\eqref{eq:Langevin} the classical stochastic force terms, $\xi_\alpha(t)$, obeying 
$\overline{\xi_\alpha(t) \xi_\beta(t')} = D_{\alpha\beta}(t)\delta(t-t')$ where the overline corresponds to the classical averaging implicit in the Langevin equation. . 

In order to determine the correlator $D_{\alpha\beta}(t)$, we average $D_{\alpha\beta}(t,t')$ over the fast degrees
of freedom corresponding to the relative time $\tau$,
\begin{equation}\label{eq:def_correlator}
D_{\alpha\beta}\left(\mathbf{X}_t\right)=\int\mathrm{d}\tau D_{\alpha\beta}\left(t+\frac{\tau}{2},t-\frac{\tau}{2}\right)\,.
\end{equation}
It is sufficient to evaluate this correlator in the fully adiabatic limit since this already ensures that the fluctuation-dissipation theorem 
be satisfied.\cite{BodePRL11,BodeBJ12} Hence we freeze the value of ${\bf X}_t$ (and by that the Hamiltonian), and evaluate the force fluctuations 
with this static Hamiltonian. Then we can
work in the frozen scattering state basis $|\psi^{{\bf X}_t\pm}\rangle$, where the Schr\"odinger force operator 
can be expressed as
\begin{equation}
\begin{split}
 \hat {\mathcal{F}}_{{\bf X}_t}
& =  - \int\frac{\mathrm{d}\varepsilon}{2\pi}\int\frac{\mathrm{d}\varepsilon'}{2\pi}a_{m}^{{\bf X}_t\dagger}(\varepsilon)\,\left[\partial_{\alpha}H_t\right]_{mk}\, a_{k}^{{\bf X}_t}(\varepsilon')\,.
\end{split} 
\label{eq:force}
\end{equation}
Here the channel summation is again implicit and $a_{m}^{{\bf X}_t\dagger}(\varepsilon)$ creates a frozen (retarded) scattering state in channel $m$ with energy $\varepsilon$ at time $t$, i.e., an eigenstate of
the frozen Hamiltonian $H_t$. 
The superscript ${\bf X}_t$ indicates the strictly adiabatic condition. Within this adiabatic approximation, changing to the Heisenberg picture simply amounts to replacing in Eq. \eqref{eq:force} the operators  $a_{k}^{{\bf X}_t}$ by the corresponding Heisenberg operators, 
$a_{k}^{{\bf X}_t}(\varepsilon) \to  a_{k}^{{\bf X}_t}(\varepsilon,t+\tau)= \mathrm{e}^{-i\,\varepsilon\, \tau }a^{{\bf X}_t}_{k}(\varepsilon)$.
\footnote{Here we set the reference time to change from Schr\"odinger to Heisenberg picture as $t$.} The correlator $D_{\alpha\beta}(t)$ can then be calculated by means of the identity \cite{ButtikerPRB92} 
\begin{equation}\label{eq:average}
\begin{split}
 \big\langle\! & a_{m}^{\dagger}(\varepsilon_1)  a_{n}(\varepsilon_2)a_{k}^{\dagger}(\varepsilon_3)a_{l}(\varepsilon_4)\! \big\rangle\! -\! \big\langle \!a_{m}^{\dagger}(\varepsilon_1)  a_{n}(\varepsilon_2)\!\big\rangle  \big\langle\! a_{k}^{\dagger}(\varepsilon_3) a_{l}(\varepsilon_4)\!\big\rangle \\
 &= (2\pi)^2\, f_m(\varepsilon_1) \, [1 \mp f_k(\varepsilon_2)] \, \delta_{m l}\,\delta_{n k} \,\delta(\varepsilon_1 - \varepsilon_4) \,\delta(\varepsilon_2 - \varepsilon_3)\,,
\end{split} 
\end{equation}
where the upper sign $(-)$ refers to fermions and the lower sign $(+)$ to bosons, and for simplicity, the label 
${\bf X}_t$ has been dropped. Applying Eq.~(\ref{eq:average}) to Eq.~(\ref{eq:def_correlator}) we finally arrive at
\begin{equation}
\begin{split}
\label{eq:correlator}
& D_{\alpha\beta}\left(\mathbf{X}_t\right) =  \int\mathrm{d}\tau\int\frac{\mathrm{d}\varepsilon}{2\pi}f_{m}(\varepsilon)\,\left[1\mp f_{k}(\varepsilon)\right]\\
&\times\left\{\big\langle\psi_{m}^{{\bf X}_t+}(\varepsilon)\big|\partial_{\alpha}H_t\big|\psi_{k}^{{\bf X}_t+}(\varepsilon)\big\rangle \big\langle\psi_{k}^{{\bf X}_t+}(\varepsilon)\big|\partial_{\beta}H_t\big|\psi_{m}^{{\bf X}_t+}(\varepsilon)\big\rangle \right\}_{s}\,,
\end{split}
\end{equation}
where $\{\ldots\}_s$ indicates symmetrization with respect to the indices $\alpha,\,\beta$. In the remainder of this section, we shall express the average adiabatic reaction force in Eq.~\eqref{eq:FAdExp} as well as the correlator in Eq.~\eqref{eq:correlator} in terms of the S and A-matrices.

\subsection{Born-Oppenheimer force}
The Born-Oppenheimer force ${\bf F}$ can be expressed solely in terms of frozen scattering states. From Eq.~\eqref{eq:FAdExp}, it is given by
\begin{equation}\label{eq:MeanFdef}
 F_{\alpha} = -\,\int \frac{\textrm{d}\varepsilon}{2\pi} \,\sum_{n} f_n(\varepsilon)\, \big\langle \psi_n^{{\bf X}_t+} (\varepsilon)\big|  \partial_\alpha  V_t \big| \psi_n^{{\bf X}_t+} (\varepsilon) \big\rangle\,.
\end{equation}
For turning Eq.~\eqref{eq:MeanFdef} into an expression involving the S-matrix, we insert a resolution of the identity $\mathbbm{1} = \int \frac{\textrm{d}\varepsilon}{2\pi} \sum_{k} \big| \psi_k^{{\bf X}_t-} (\varepsilon) \big\rangle \big\langle \psi_k^{{\bf X}_t-} (\varepsilon)\big|$, 
\begin{equation}
\begin{split}
 F_{\alpha} = -\int \frac{\textrm{d}\varepsilon}{2\pi}\int \frac{\textrm{d}\varepsilon'}{2\pi} \sum_{nm} &f_n(\varepsilon)\, \big\langle \psi_n^{{\bf X}_t+} (\varepsilon)\big| \psi_m^{{\bf X}_t-} (\varepsilon') \big\rangle\\
& \times\big\langle \psi_m^{{\bf X}_t-} (\varepsilon')\big| \partial_\alpha  V_t \big| \psi_n^{{\bf X}_t+} (\varepsilon) \big\rangle\, .
\end{split}
\end{equation}
Since the dependence on time $t$ is parametric through $\mathbf{X}_t$, from Eq.~\eqref{eq:dtS} we have
\begin{equation}
  \partial_\alpha S_{n k} (\varepsilon) = - i \, \big\langle \psi_n^{{\bf X}_t-}(\varepsilon)\big|\partial_\alpha V\big| \psi_k^{{\bf X}_t+}(\varepsilon) \big\rangle \,.
\end{equation}
Putting this together with the expression for the frozen S-matrix Eq.~\eqref{eq:FrozenS} we obtain
\begin{equation}
\begin{split}
 F_{\alpha} = \int \frac{\textrm{d}\varepsilon}{2\pi i} \sum_{nm} &f_n(\varepsilon)S_t^{\dagger nm}(\varepsilon)\partial_\alpha S_t^{mn}(\varepsilon)\,.
\end{split}
\end{equation}
In matrix notation, this gives the Born-Oppenheimer force 
\begin{equation}\label{eq:Born-Oppenheimer_force}
 F_{\alpha}(\mathbf{X}_t) = \int \frac{\textrm{d}\varepsilon}{2\pi i} \sum_n f_n(\varepsilon)\textrm{tr}\left\{\Pi_n S_t^{\dagger}(\varepsilon)\partial_\alpha S_t(\varepsilon) \right\}\,,
\end{equation}
where $\textrm{tr}\{\ldots\} $ denotes a trace over scattering channels, and $\Pi_n$ is a projector onto channel $n$. Eq.~(\ref{eq:Born-Oppenheimer_force}) coincides with the one obtained in Ref.~\onlinecite{BodePRL11} {\it via} a non-equilibrium Keldysh calculation for the current-induced-forces in a nanoelectromechanical system.

The expression given in Eq. \eqref{eq:Born-Oppenheimer_force} can be motivated by realizing its connection with the Friedel sum rule. \cite{FriedelPM52} Considering a finite system with discrete energy levels $E_t^i=E^i(\mathbf{X}_t)$, the Born-Oppenheimer force in equilibrium is given by
\begin{equation}\label{eq:BOdiscrete}
F_{\alpha}(\mathbf{X}_t) =-\sum_i f\left(E_t^i\right) \partial_\alpha E_t^i\,
\end{equation}
If we replace $E_t^i \rightarrow \int\textrm{d}\,\varepsilon\,\varepsilon\delta(\varepsilon-E_t^i)$ in Eq. \eqref{eq:BOdiscrete}, we can take the limit of the system size to infinity by writing the number of states up to energy $\varepsilon$ as $N(\varepsilon,\mathbf{X}_t)=\int_{-\infty}^\varepsilon \mathrm{d}\varepsilon'\nu(\varepsilon',\mathbf{X}_t)$ where $\nu$ is the density of states and we have used the identity $\partial_\alpha \Theta\left(\varepsilon-E_t^i\right)=-\delta\left(\varepsilon-E_t^i\right)\,\partial_\alpha E_t^i$. In this limit, Eq. \eqref{eq:BOdiscrete} takes the form
\begin{equation}
 F_{\alpha}(\mathbf{X}_t)=\int \mathrm{d} \varepsilon \,f(\varepsilon)\partial_\alpha N(\varepsilon,\mathbf{X}_t)\,.
\end{equation}
The quantity $\partial_\alpha N$ is known as the emissivity \cite{BuettikerZPB94} and plays a key role in the problem of adiabatic quantum pumping. \cite{BrouwerPRBR98} [Note that expressions of the type $S_t^{\dagger} \partial_\alpha S_t$ also appear in the context of quantum pumping as ``response matrices''.] Making use of the general expression for the Friedel sum rule in terms of S-matrices, \cite{LangerPR61}
\begin{equation}
N (\varepsilon,\mathbf{X}_t)=\frac{1}{2\pi i}\textrm{tr}\left\{\ln S_t(\varepsilon)\right\}\,,
\end{equation}
the emissivity can be expressed as
\begin{equation}
\partial_\alpha N (\varepsilon,\mathbf{X}_t)=\frac{1}{2\pi i}\textrm{tr}\left\{ S_t^{\dagger}(\varepsilon)\partial_\alpha S_t(\varepsilon) \right\}\,,
\end{equation}
and hence we recover Eq. \eqref{eq:Born-Oppenheimer_force} for the particular case that the system is in equilibrium.

\subsection{Friction and geometric magnetic force}
We now express the tensor $\boldsymbol\gamma$ in terms of the frozen S-matrix and the first order non-adiabatic correction, the A-matrix. The first order correction to the Born-Oppenheimer force is given by the two last lines of Eq.~\eqref{eq:FAdExp}. With the aid of the chain rule $\partial_t= \dot{X}_\alpha\partial_\alpha$, it is straightforward to show that
\begin{widetext}
 \begin{equation}\label{eq:gamma}
\begin{split}
\gamma_{\alpha\beta}&=i\int\frac{\mathrm{d}\varepsilon}{2\pi}f_k(\varepsilon)\big\langle \psi_k^{{\bf X}_t+}\big|\partial_{\beta}V_t\left(G^{A}_t\right)^{2}\partial_{\alpha}V_t\big|\psi_k^{{\bf X}_t+}\big\rangle-i\int\frac{\mathrm{d}\varepsilon}{2\pi}f_k(\varepsilon)\big\langle \psi_k^{{\bf X}_t+}\big|\partial_{\alpha}V_t\left(G^{R}_t\right)^{2}\partial_{\beta}V_t\big|\psi_k^{{\bf X}_t+}\big\rangle\,,
\end{split}
\end{equation}
where we have omitted energy variables and left the sum over $k$ implicit. We split the tensor $\boldsymbol\gamma$ into a symmetric part $\gamma_{\alpha \beta}^s = 1/2 \, (\gamma_{\alpha \beta} + \gamma_{\beta \alpha})$, corresponding to the friction force, and an antisymmetric part $\gamma_{\alpha \beta}^a = 1/2 \, (\gamma_{\alpha \beta} - \gamma_{\beta \alpha})$, corresponding to the emergent Lorentz force.

We first consider the symmetric, dissipative contribution
\begin{equation}\label{eq:DefGammaS}
\begin{split}
\gamma_{\alpha\beta}^{s}&=i\int\frac{\mathrm{d}\varepsilon}{2\pi}f_k(\varepsilon)\big\langle \psi_k^{{\bf X}_t+}\big|\partial_{\alpha}V_t\left(G^{A}_t\right)^{2} \partial_{\beta}V_t\big|\psi_k^{{\bf X}_t+}\big\rangle _{s}-i\int\frac{\mathrm{d}\varepsilon}{2\pi}f_k(\varepsilon)\big\langle\psi_k^{{\bf X}_t+}\big|\partial_{\alpha}V_t\left(G^{R}_t\right)^{2} \partial_{\beta}V_t\big|\psi_k^{{\bf X}_t+}\big\rangle _{s}\,,
\end{split}
\end{equation}
 Using the identity $\left(G^{A}_t\right)^{2}-\left(G^{R}_t\right)^{2}=-\partial_{\varepsilon}(G^{A}_t-G^{R}_t)=-2\pi i\partial_{\varepsilon}\delta(\varepsilon-H_t)$ and integrating by parts we obtain
\begin{equation}
\begin{split}
 \gamma_{\alpha\beta}^{s}&=\int\mathrm{d}\varepsilon\left[-\partial_{\varepsilon}f_k(\varepsilon)\right]\big\langle \psi_k^{{\bf X}_t+}\big|\partial_{\alpha}V_t\delta(\varepsilon-H_t)\partial_{\beta}V_t\big|\psi_k^{{\bf X}_t+}\big\rangle _{s}  -\int\mathrm{d}\varepsilon f_k(\varepsilon)\big\langle \partial_{\varepsilon}\psi_k^{{\bf X}_t+}\big|\partial_{\alpha}V_t\delta(\varepsilon-H_t)\partial_{\beta}V_t\big|\psi_k^{{\bf X}_t+}\big\rangle _{s}\\
 &   -\int\mathrm{d}\varepsilon f_k(\varepsilon)\big\langle \psi_k^{{\bf X}_t+}\big|\partial_{\alpha}V_t\delta(\varepsilon-H_t)\partial_{\beta}V_t\big|\partial_{\varepsilon}\psi_k^{{\bf X}_t+}\big\rangle _{s}\,.
\end{split}
\end{equation}
We now insert a resolution of the identity $\sum_l\int\frac{\mathrm{d}\varepsilon'}{2\pi}|\psi_l^{{\bf X}_t-}(\varepsilon')\rangle\langle\psi_l^{{\bf X}_t-}(\varepsilon')|$ between the two potential terms and find
\begin{equation}
\begin{split}\label{eq:auxaux}
\gamma_{\alpha\beta}^s & =  -\int\frac{\mathrm{d}\varepsilon}{2\pi}\partial_{\varepsilon}f_k(\varepsilon)\left\{\big\langle \psi_k^{{\bf X}_t+}\big|\partial_{\alpha}V_t\big|\psi_l^{{\bf X}_t-}\big\rangle \big\langle \psi_l^{{\bf X}_t-}\big|\partial_{\beta}V_t\big|\psi_k^{{\bf X}_t+}\big\rangle \right\}_{s}\\
&  -\int\frac{\mathrm{d}\varepsilon}{2\pi}f_k(\varepsilon)\left\{\big\langle \partial_{\varepsilon}\psi_k^{{\bf X}_t+}\big|\partial_{\alpha}V_t\big|\psi_l^{{\bf X}_t-}\big\rangle \big\langle \psi_l^{{\bf X}_t-}\big|\partial_{\beta}V_t\big|\psi_k^{{\bf X}_t+}\big\rangle \right\}_{s}\\
 &   -\int\frac{\mathrm{d}\varepsilon}{2\pi}f_k(\varepsilon)\left\{\big\langle \psi_k^{{\bf X}_t+}\big|\partial_{\alpha}V_t\big|\psi_l^{{\bf X}_t-}\big\rangle \big\langle \psi_l^{{\bf X}_t-}\big|\partial_{\beta}V_t\big|\partial_{\varepsilon}\psi_k^{{\bf X}_t+}\big\rangle \right\}_{s}\,.
\end{split}
\end{equation}
Comparing this expression with Eqs.~\eqref{eq:A2} and \eqref{eq:dtS} and using the definition of the A-matrix given in Eq.~\eqref{eq:Aalpha}, leads immediately to the final result
\begin{equation}\label{eq:sym_friction}
\begin{split}
 \gamma_{\alpha \beta}^s (\mathbf{X}_t) &= \int\frac{\textrm{d}\varepsilon}{4\pi}\sum_n [-\partial_\varepsilon f_n(\varepsilon)] \textrm{tr}\left\{   \Pi_{\substack{n}} \partial_\alpha S_t^\dagger(\varepsilon)\partial_\beta S_t(\varepsilon) \right\}_s + \int\frac{\textrm{d}\varepsilon}{2 \pi i} \sum_n f_n(\varepsilon)\textrm{tr}\left\{   \Pi_{\substack{n}}  \left[  \partial_\alpha S_t^\dagger(\varepsilon)  A_t^\beta(\varepsilon) - A_t^{\beta\dagger}(\varepsilon) \partial_\alpha S_t(\varepsilon)    \right]     \right\}_s\,.
\end{split}
\end{equation}
\end{widetext}
Equation~\eqref{eq:sym_friction} recovers the frictional force obtained first in Ref.~\onlinecite{BodePRL11}. Thus, we conclude that the classical degrees of freedom are indeed subject to a friction force due to the coupling to a quantum mechanical scattering system. This is in stark contrast with the coupling to a finite quantum system where Berry and Robbins find that the frictional contribution to the adiabatic reaction force vanishes.\cite{BerryPRSL93}

The antisymmetric part of the damping matrix has the role of an effective
orbital magnetic field acting on the multidimensional space of $\mathbf{X}$. From Eq.~\eqref{eq:gamma}, it is given by 
\begin{equation}
\begin{split}
\gamma_{\alpha\beta}^{a}&=\int\frac{\mathrm{d}\varepsilon}{2\pi i}f(\varepsilon)\big\langle \psi^{{\bf X}_t+}\big|\partial_{\alpha}V_t\left(G^{A}_t\right)^{2}\partial_{\beta}V_t\big|\psi^{{\bf X}_t+}\big\rangle_{a}\\
&+\int\frac{\mathrm{d}\varepsilon}{2\pi i}f(\varepsilon)\big\langle \psi^{{\bf X}_t+}\big|\partial_{\alpha}V_t\left(G^{R}_t\right)^{2} \partial_{\beta}V_t\big|\psi^{{\bf X}_t+}\big\rangle_{a}\,.
\end{split}
\end{equation}
To evaluate this expression we observe that, using Eq.~\eqref{eq:A2} and $(G^R_t)^2=-\partial_\varepsilon G^R_t$, it follows straightforwardly that
\begin{equation}
\left\{\partial_{\alpha}A_t^{\beta}\right\}_{a}  =  \big\langle \psi^{{\bf X}_t-}\big|\partial_{\beta}V_t\left(G^{R}_t\right)^{2}\partial_{\alpha}V_t\big|\psi^{{\bf X}_t+}\big\rangle_{a}  \,,
\end{equation}
with an analogous expression involving $G^A_t$. Similar manipulations to the ones employed in Eq.~\eqref{eq:auxaux} lead to the result for the Lorentz-like term of the ${\boldsymbol\gamma}$ matrix
\begin{equation}\label{eq:antisym_friction}
\begin{split}
 \gamma_{\alpha \beta}^a(\mathbf{X}_t) =& \int\frac{\textrm{d}\varepsilon}{2 \pi i} \,\sum_{\substack{n}}\,f_n(\varepsilon)\\
 &\times\textrm{tr} \left\{  \Pi_n \left[ S_t^\dagger(\varepsilon)\partial_\beta A_t^\alpha(\varepsilon)  - \partial_\beta A_t^{\alpha\dagger} (\varepsilon) S_t^{\vphantom{\dagger}}(\varepsilon) \right] \right\}_a  \,. 
\end{split}
\end{equation}
This expression agrees with the one obtained in Ref.~\onlinecite{BodePRL11}. Note that within the scattering formalism all the above relations are well defined and we do not encounter any divergences in contrast
to Ref.~\onlinecite{LueNanoLett10}. The reason for this is that the particles spend a finite time in the scattering region, which is implicit in the time-dependent scattering formalism. 

\subsection{Stochastic force}

The stochastic force can be written in terms of the frozen S-matrix by inserting two resolutions of the identity of the form
$\mathbbm{1} = \int \frac{\textrm{d}\varepsilon}{2\pi} \sum_{k} | \psi_k^{{\bf X}_t-} (\varepsilon) \rangle \langle \psi_k^{{\bf X}_t-} (\varepsilon)|$ into Eq.~\eqref{eq:correlator}. We can then identify the frozen S-matrix by use of Eq.~\eqref{eq:FrozenS} as well as its derivatives given by Eq.~\eqref{eq:STmat}. This yields
\begin{equation}
\begin{split}
 D_{\alpha\beta}  &= \int \frac{\textrm{d}\varepsilon}{2\pi}   f_n(\varepsilon)\, \left[1 \mp f_m(\varepsilon)\right] \\
 &  \times\left\{ \vphantom{\int} \partial_\alpha S_t^{\dagger n k}(\varepsilon) \, S_t^{k m}(\varepsilon) \, S_t^{\dagger m l}(\varepsilon) \, \partial_\beta S_t^{l n}(\varepsilon) \right\}_s\,.
\end{split}
\end{equation}
In matrix notation, this can equivalently be written as
\begin{equation}
\begin{split}
\label{eq:result_correlator}
D_{\alpha \beta}(\mathbf{X}_t)& =    \sum_{n m}\, \int \frac{\textrm{d}\varepsilon}{2 \pi}   \,f_n(\varepsilon)\,\left[1 \mp f_m(\varepsilon)\right] \\
 & \times\textrm{tr}\left\{  \Pi_n \,\left[ S_t^{\dagger}(\varepsilon) \, \partial_\alpha S_t(\varepsilon)  \right]^\dagger\,\Pi_m\, S_t^{\dagger}(\varepsilon)\, \partial_\beta S_t^{\vphantom{dagger}}(\varepsilon)   \right\}_s\,,
\end{split}
\end{equation}
which agrees with the expression in Ref.~\onlinecite{BodePRL11} when dealing with fermions. It can be shown that this expression fulfills the fluctuation-dissipation theorem in equilibrium,\cite{BodePRL11} ${\bf D}=2T{\boldsymbol \gamma}_s$, with ${\boldsymbol\gamma}_s$ given by Eq.~\eqref{eq:sym_friction} and evaluated in equilibrium. 


\section{Conclusion}
\label{sec:Concl}
Slow degrees of freedom coupled to a fast quantum system are subject to adiabatic reaction forces. Currently, these forces play a pivotal role in the context of nanoelectromechanical and spintronics systems with a slow mechanical mode or spin coupled to fast electronic  degrees of freedom. While early work on the adiabatic reaction forces focused on closed quantum systems, nanoelectromechanicals and spintronics typically involve open electronic systems  driven out of equilibrium by voltage sources.

These developments have motivated us to consider adiabatic reaction forces for a generic model of a slow classical degree of freedom coupled to a quantum mechanical scattering system. Non-equilibrium is incorporated into this (non-interacting) many-body model by assuming that the filling of the incoming scattering channels is controlled by various reservoirs. In the context of nanoelectromechanics and spintronics, this model follows naturally if the electronic degrees of freedom take the form of a mesoscopic Landauer-B\"uttiker conductor.

It was shown recently within a Keldysh Green's function approach \cite{BodePRL11,BodeBJ12} that the adiabatic reaction forces can be expressed entirely in terms of the adiabatic S-matrix and its first non-adiabatic correction, the A-matrix. The main result of the present paper is an alternative derivation of these results within the setting and with the methods of scattering theory.

In addition to being more natural and more direct, this derivation has several further advantages. To start with, we present useful expressions for the A-matrix in terms of the adiabatic scattering states which should simplify its calculation for specific applications. The general setting within the context of scattering theory
facilitates comparison with the earlier results on adiabatic reaction forces for closed quantum systems. Most prominently, there is no frictional force for closed quantum systems while such a force emerges naturally for a quantum-mechanical scattering system. Moreover, the approach clarifies the limits of validity. While for closed quantum systems, the adiabatic condition involves the level spacing, the latter is replaced here by the dwell time of the fast system in the scattering region. Finally, the general setting  emphasizes the generality and wide applicability of our results. The fast quantum system can be fermionic as in nanoelectromechanics, a spin degree of freedom as in spintronics, or bosonic as in optomechanics or cold-atom systems.

\section*{Acknowledgments}
We would like to acknowledge useful discussions with N. Bode, P. Brouwer, O. Entin-Wohlman and B. Rosenow, as well as financial support by the EU-NKTH GEOMDISS project and the Deutsche Forschungsgemeinschaft through Sfb 658, SPP 1459 as well as a Mercator Professorship. 
\begin{appendices}
 \numberwithin{equation}{section} 
\section{Relation between the S and A-Matrix}\label{sec:AandS-matrix}
Here we prove the identity given in Eq.~(\ref{eq:S_and_A}). For better readability we omit the channel index and energy dependence (all quantities are evaluated at the same energy $\varepsilon$). Starting with Eq.~\eqref{eq:A2} we find
\begin{equation}
\begin{split}\label{eq:step}
& S_t^{\dagger}A_t+A_t^{\dagger}S_t \\
&=  -\partial_{\varepsilon}\big\langle \psi^{{\bf X}_t+}\big|\dot{V}_t\big|\psi^{{\bf X}_t+}\big\rangle+\frac{i}{2}\left[S_t^{\dagger}\partial_{\varepsilon}\partial_{t}S_t-\partial_{\varepsilon}\partial_{t}S_t^{\dagger}S_t\right]\\
 & =  -\frac{i}{2}\partial_{\varepsilon}\left[S_t^{\dagger}\partial_t S_t-\partial_t S_t^{\dagger}S_t\right]+\frac{i}{2}\left[S_t^{\dagger}\partial_{\varepsilon}\partial_{t}S_t-\partial_{\varepsilon}\partial_{t}S_t^{\dagger}S_t\right]\\
 & =  \frac{i}{2}\left[\partial_{t}S_t^{\dagger}\partial_{\varepsilon}S_t-\partial_{\varepsilon}S_t^{\dagger}\partial_{t}S_t\right]\,
\end{split}
\end{equation}
where we made use of Eq.~\eqref{eq:dtS} to obtain the second equality in Eq.~\eqref{eq:step}.
\section{Application: A quantum dot coupled to leads}
\label{sec:Application}
As we mentioned above, the S and A-matrix expressions presented in this work for the reaction forces were obtained first, within a different formalism, in Ref.~\onlinecite{BodePRL11} for the forces that an applied current exerts over the slow vibrational degrees of freedom of a nanomechanical oscillator. In this section we show how our formalism relates to the one presented in Ref.~\onlinecite{BodePRL11} where the Hamiltonian
\begin{equation}\label{eq:Niels_Hamiltonian}
H(\mathbf{X})= H_X + H_L + H_D + H_T 
\end{equation}
was considered, which models a quantum dot connected to leads. The ``heavy'' classical degrees of freedom $\mathbf{X}(t) = \{X_1(t),X_2(t) \ldots X_N(t)\}$ in this case are the mechanical vibrational modes of the dot, which couple to the electrons in the dot. The different terms of the Hamiltonian in Eq.~\eqref{eq:Niels_Hamiltonian} are given by 
\begin{align}
H_L&= \int \frac{\textrm{d}\varepsilon}{2\pi}\sum_{\eta}\left(\varepsilon-\mu_{\eta}\right)c_{\eta}^{\dagger}(\varepsilon)c_{\eta}(\varepsilon) \\
H_X &= \sum_\nu \left[ \frac{P_\nu^2}{2M_\nu} + U(\mathbf{X})  \right] \\
H_D &= \sum_{mm'}d_{m}^{\dagger}\left[h(\mathbf{X})\right]_{mm'}d_{m'} \\
H_T &= \int \frac{\textrm{d}\varepsilon}{\sqrt{2\pi}}\sum_{\eta m}\left(c_{\eta}^{\dagger}(\varepsilon)W_{\eta m}(\varepsilon)d_{m}+h.c.\right) \,.
\end{align}
$H_L$ models the leads, where $c_{\eta}^{\dagger}(\varepsilon)$ [$c_{\eta}(\varepsilon)$] creates [annihilates] electrons in a flux normalized state $|\phi_\eta(\varepsilon)\rangle$ incoming from $\eta$ with chemical potential $\mu_\eta$ ($\eta$ combines channel and lead index, the chemical potential depends only on the lead index). $H_X$ represents the free evolution of the mechanical degrees of freedom of the dot. $H_D$ is the Hamiltonian of the dot, containing the electronic levels plus the coupling of the electrons in the dot to $\mathbf{X}$ via a general function $h(\mathbf{X})$. The operators $d_{m}^{\dagger}$ ($d_{m}$) create (annihilate) a dot-electron in the state $|m\rangle$. Finally, $H_T$ indicates the tunneling process between the dot's levels and the leads with tunneling amplitude $W_{\eta m}(\varepsilon)=\langle \phi_\eta(\varepsilon)|W|m\rangle/\sqrt{2\pi}$.

The electronic part of the Hamiltonian in Eq.~(\ref{eq:Niels_Hamiltonian}) can be interpreted as a scattering problem, where the free Hamiltonian is given by $H_L$ and the dot defines a scattering potential $V = \Pi_D W^\dagger \Pi_L + \Pi_L W \Pi_D + \Pi_D H_D\Pi_D$ where $\Pi_L$ and $\Pi_D$ project onto the lead and dot space, respectively. (Note that $\Pi_L \cdot \Pi_D = \Pi_D \cdot \Pi_L = 0$.)

We can then write the Lippmann-Schwinger equation as
\begin{align}\label{eq:Lipp_Schw}
 \big| \psi_\eta^{{\bf X}_t+}(\varepsilon) \big\rangle = \Pi_L \, \big| \phi_\eta(\varepsilon) \big\rangle + G^{R}_t(\varepsilon)V_t \, \Pi_L \, \big|\phi_\eta(\varepsilon) \big\rangle
\end{align}
where $G^{R}_t(\varepsilon) = (\varepsilon - H _t+ i\eta)^{-1}$ is the frozen Green's function of the dot plus lead. The projection of Eq.~(\ref{eq:Lipp_Schw}) onto the dot space takes the form
\begin{align}
  \Pi_D\,\big| \psi_\eta^{{\bf X}_t+}(\varepsilon) \big\rangle &= \Pi_D \,G^{R}_t(\varepsilon) \,\Pi_D \, W^\dagger \, \Pi_L \, \big| \phi_\eta(\varepsilon) \big\rangle \notag\\
  &= G_D^{R}(\varepsilon) \, W^\dagger \,\big|\phi_\eta(\varepsilon)\big\rangle \label{eq:Lipp_dot}
\end{align}
with $G_D^{R}(\varepsilon) = \Pi_D \,G^{R}_t(\varepsilon) \,\Pi_D$, the dot's frozen Green function. 

We are interested in an explicit expression for the A-matrix. Since $\partial_t{V}_t = \Pi_D \partial_t{H}_D \Pi_D$, we obtain from Eq.~(\ref{eq:A}) 
\begin{align}
 A_t^{\eta \mu}(\varepsilon)&= \frac{1}{2} \left[\big\langle  \partial_{\varepsilon}  \psi_\eta^{{\bf X}_t-}(\varepsilon)\big| \Pi_D   \partial_t{H}_D \, \Pi_D \, \big|\psi_\mu^{{\bf X}_t+}(\varepsilon)  \big\rangle   \right. \notag\\
 &\quad \left.-\big\langle \psi_\eta^{{\bf X}_t-}(\varepsilon)\big| \, \Pi_D \, \partial_t{H}_D \,   \Pi_D\, \big|\partial_{\varepsilon}\psi_\mu^{{\bf X}_t+}(\varepsilon) \big\rangle \right].
\end{align}
Using Eq.~\eqref{eq:Lipp_dot} this can be cast into the form
\begin{equation}\label{eq:bode}
\begin{split}
A_t^{\eta\mu} & = \pi \left\{\partial_{\varepsilon}\left(W_{\eta k}\left[G_{D}^{R}\right]_{kl}\right)\left[\partial_t{h}_{lm}\right]\left[G_{D}^{R}\right]_{mn} W^{\dagger}_{n \mu}\right. \\ 
& -\left.W_{\eta k}\left[G_{D}^{R}\right]_{kl}\left[ \partial_t{h}_{lm}\right] \partial_{\varepsilon}\left(\left[G_{D}^{R}\right]_{mn}W^{\dagger}_{n\mu}\right)\right\}\,,
\end{split}
\end{equation}
where the summation over repeated indices is implied. Applying our formalism to the Hamiltonian \eqref{eq:Niels_Hamiltonian} therefore indeed recovers the expression of the A-matrix given by Ref.~\onlinecite{BodePRL11}.
\end{appendices}

\bibliographystyle{apsrev}
\bibliography{LongCIF_Silvia}
\end{document}